\documentclass[12pt,a4paper]{article}

\usepackage{graphicx}
\usepackage{amssymb} 
\usepackage{amsthm} 
\usepackage{lineno}
\usepackage{subfigure}
\usepackage[babel=true]{csquotes}    

\def\beq     {\begin{equation}}
\def\eeq     {\end{equation}}

\newcommand{\ie}{{\it i.e.}}
\newcommand{\eg}{{\it e.g.}}
\newcommand{\etal}{{\it et al.}}

\def\lsim{\raise0.3ex\hbox{$<$\kern-0.75em\raise-1.1ex\hbox{$\sim$}}}
\def\gsim{\raise0.3ex\hbox{$>$\kern-0.75em\raise-1.1ex\hbox{$\sim$}}}

\def\xF    {\mbox{$x_F$}}

\usepackage{ifpdf}
\ifpdf
\usepackage[pdftex]{hyperref}
\else
\usepackage[hypertex]{hyperref}
\fi

\hypersetup{
  pdftitle={},%
  pdfauthor={},%
  pdfsubject={},%
  pdfkeywords={},%
  pdfstartview={},%
  bookmarksopen=true, breaklinks=true, debug=true, %
  colorlinks=true, linkcolor=red, citecolor=blue, urlcolor=blue
}

\begin{document}


\textwidth=135mm
 \textheight=200mm
\begin{center}

{\bfseries Spin physics at A Fixed-Target ExpeRiment at the LHC (AFTER$@$LHC)\footnote{{\small Talk at the SPIN 2012 Conference, JINR, Dubna, September 17 - 22,
2012.}}}
\vskip 5mm

A.~Rakotozafindrabe$^a$,
M.~Anselmino$^b$,
R.~Arnaldi$^b$,
S.J.~Brodsky$^c$,
V.~Chambert$^d$,
J.P. Didelez$^d$,
E.G.~Ferreiro$^e$,
F.~Fleuret$^f$,
B. Genolini$^d$,
C.~Hadjidakis$^d$,
J.P.~Lansberg$^d$,
C.~Lorc\'e$^d$,
P. Rosier$^d$,
I.~Schienbein$^g$,
E.~Scomparin$^b$,
and {U.I.~Uggerh\o j}$^h$
\vskip 5mm

{\small 
$^a${\it IRFU/SPhN, CEA Saclay, 91191 Gif-sur-Yvette Cedex, France}\\
$^b${\it INFN Sez. Torino, Via P. Giuria 1, I-10125, Torino, Italy}\\ 
$^c${\it SLAC National\,Accelerator\,Laboratory, Stanford U., Menlo Park, CA 94025,USA}\\
$^d${\it IPNO, Universit\'e Paris-Sud, CNRS/IN2P3, F-91406, Orsay, France}\\
$^e${\it Dept. de F{\'\i}sica de Part{\'\i}culas, USC, 15782 Santiago de Compostella, Spain}\\
$^f${\it LLR, \'Ecole Polytechnique, CNRS/IN2P3,  91128 Palaiseau, France}\\
$^g${\it LPSC, Univ. Joseph Fourier, CNRS/IN2P3/INPG, 38026 Grenoble, France}\\
$^h${\it Department of Physics and Astronomy, University of Aarhus, Denmark}  
}
\\
\end{center}
\vskip 5mm

\centerline{\bf Abstract}
We outline the opportunities for spin physics which are offered by a next generation 
and multi-purpose fixed-target experiment exploiting the proton LHC beam extracted by a bent crystal. 
In particular, we focus on the study of single transverse spin asymetries with the polarisation
of the target.
\vskip 10mm


\section{Introduction}

Fixed-target experiments lead to numerous breakthroughs in particle and nuclear physics. In particular
they contributed to the discovery of anomalously large single~\cite{Adams:1991cs} and double-spin~\cite{Crosbie:1980tb} 
correlations in hadron-hadron collisions.

New opportunities~\cite{Brodsky:2012vg,Lansberg:2012kf} are at reach thanks to the LHC beam of 7~TeV protons interacting on a fixed target, 
be it polarised or unpolarised. Such opportunities can be studied by a future multi-purpose experiment, 
thereafter named AFTER for \enquote{A Fixed-Target ExperRiment}. The LHC proton beam colliding on 
fixed targets releases a cms energy close to 115 GeV, \ie\ an energy never reached in a 
fixed-target experiment, between SPS and RHIC. 

One of the essential advantages of a fixed-target experiment is the possibility to polarize the 
target (see \eg~\cite{Crabb:1997cy}) to measure Single Spin Asymmetry (SSA) in a number of hard reactions.
We discuss here the opportunities offered by the measurements of such SSAs.

\section{Beam extraction and target polarisation}

The extraction of beams by means of bent-crystal chanelling 
offers an ideal and cost-effective way to obtain a clean and very 
collimated beam without specific limitation on its energy. This would not alter  the LHC beam 
performance~\cite{LUA9,Uggerhoj:2005xz}. 
The \enquote{smart collimator} solution on the 7-TeV LHC beam will be tested by 
the CERN LUA9 collaboration after the 2013 shutdown~\cite{LHCC107}. 
Another proposal, which deserves to be further investigated, is to \enquote{replace} the kicker-modules in LHC section IR6 
by a bent crystal~\cite{Uggerhoj:2005xz}.

A significant fraction of the beam loss can then been extracted, with an intensity reaching
$5\times 10^8$ $p^+$s$^{-1}$. This corresponds to an average extraction  per bunch per revolution of mini-bunches
of about 15 $p^+$. With typical targets, the yearly ($10^7$ s) integrated luminosities are of the 
order of an inverse~fb. For more details, the reader is guided to~\cite{Brodsky:2012vg}.

Whereas outstanding luminosities can be obtained, the intensity of the extracted beam 
is not extremely large and  does not constrain the choice of the target polarisation technique.
Since the beam is highly energetic, one expects a minimum ionisation and a low heating of the target. 
The heating power due to the AFTER beam would be of the order of 50 $\mu\hbox{W}$ for a typical 
1~cm thick target. This allows one  to maintain target temperatures as low as 50~mK. Relaxation times 
can last as long as one month in the spin-frozen mode.
 As regards the damages on the target, they typically  arise after an irradiation of $10^{15} p^+ \hbox{cm}^{-2}$~\cite{Baur00,meyer}, 
\ie~one month of beam in our case.

Yet, the available space in the underground LHC complex can be a major constraint. 
This restricts the choice  to 
polarisation by continuous {\it Dynamic Nuclear Polarisation} DNP or to a HD target~\cite{didelez}. 
Both take less space than the frozen-spin 
machinery. 
CERN has a long tradition of DNP for various materials such as NH$_3$, Li$_6$D~\cite{berlin}.  
Experts of DNP can still be found worlwide, whereas 
HD target makers are more rare. One can cite two 
groups: one at TJNAF (USA) and the other at RCNP (Japan)~\cite{kohri}.  We are hopeful that 
AFTER will motivate
our colleagues working on DNP to revisit  the necessary technology~\cite{solem}.

\section{Physics}

It has recently been re-emphasized that a type of parton distribution functions,  the 
"Sivers functions"~\cite{Sivers:1989cc}, can be accessed in SSAs for hard reactions 
which involve a transversely polarized proton (see \eg~\cite{Brodsky:2002cx}). 
The Sivers functions encode information on the correlation between the proton spin and
the momentum inside the proton of a parton in the transverse direction. 

As such 
they are linked to the orbital motion of partons in the polarised nucleon. Such
asymmetries have been observed in hadron-hadron collisions in forward pion~\cite{Adams:1991cs,Arsene:2008mi,Abelev:2008qb} and 
kaon production~\cite{Arsene:2008mi} at Fermilab and Brookhaven.
The high energy of the LHC beam combined 
with a well designed detector will open a unique access to the large negative-\xF\ domain, 
essentially unexplored. In this region, the SSA are sensitive to partons with large momentum fractions
in the polarised nucleon, $x^\uparrow$. This is where one expects the largest correlation between
the motion of the parton and the spin of the nucleon.

Yet, the Sivers effect for gluons is relatively -- if not completely -- unknown. This calls for the study of
SSAs with gluon sensitive reactions, such as prompt photons, open and closed heavy flavours.
Measurements of SSAs in $p^\uparrow p \to \gamma X$ was suggested more than ten years ago in~\cite{Qiu:1991wg}
and was shown to be sensitive to the gluon Sivers function~\cite{Schmidt:2005gv} for $\sqrt{s}$ of the
order of 100 GeV, where the asymmetry may be as large as 10\%~\cite{D'Alesio:2006fp}.
Further suggestions to look at gluon-sensitive SSAs were expressed: to look 
for SSA in photon-jet production~\cite{Bacchetta:2007sz}
with a constraint on the pseudo-rapidities of both
the photon and the jet and to look at SSAs in photon-pair production~\cite{Qiu:2011ai}.

We also guide the reader to two recents publication on spin physics with AFTER, 
following our first publications on AFTER~\cite{Brodsky:2012vg,Lansberg:2012kf}. The first is
about SSA in Drell-Yan reactions~\cite{Liu:2012vn} as a way to further constrain the 
quark Sivers effect. The second is about the production of spinless
quarkonium in unpolarised proton-proton collisions~\cite{Boer:2012bt}  as a way to
probe the dynamics of linearly polarized gluons inside unpolarized protons.

\section{Conclusion}

A fixed-target experiment using the multi-TeV proton beam of the LHC extracted by a bent crystal offers
outstanding opportunities for spin physics  at unprecedented laboratory energies and momentum transfers.
As we mentioned, the target polarisation at AFTER
should not cause any specific difficulties. Thanks to the extremely high luminosity of the fixed target
mode -- above the inverse femtobarn per year -- , single-spin asymetries could then be measured 
with high accuracy, in particular for gluon sensitive probes.
This would open the path for systematic studies of the contribution of the gluon motion to the nucleon spin.






\begin{thebibliography}{99}


\vspace*{-0.15cm}\bibitem{Adams:1991cs}
  D.~L.~Adams \etal\  [E704 Collaboration],
  Phys.\ Lett.\ B {\bf 264} (1991) 462.

\vspace*{-0.15cm} \bibitem{Crosbie:1980tb}
  E.~A.~Crosbie \etal, 
  Phys.\ Rev.\ D {\bf 23} (1981) 600.

\vspace*{-0.15cm} \bibitem{Brodsky:2012vg}
  S.~J.~Brodsky, F.~Fleuret, C.~Hadjidakis, J.~P.~Lansberg,
Phys.\ Rept.\ {\bf 522} (2013) 239.

\vspace*{-0.15cm} \bibitem{Lansberg:2012kf}
  J.~P.~Lansberg, S.~J.~Brodsky, F.~Fleuret and C.~Hadjidakis,
Few Body Syst.\  {\bf 53} (2012) 11.


\bibitem{Crabb:1997cy}
  D.~G.~Crabb, W.~Meyer,
  Ann.\ Rev.\ Nucl.\ Part.\ Sci.\  {\bf 47} (1997) 67.

\vspace*{-0.15cm} \bibitem{LUA9} W.~Scandale \etal\ [LUA9], CERN-LHCC-2011-007, 2011.

\vspace*{-0.15cm} \bibitem{Uggerhoj:2005xz}
  E.~Uggerh\o j, U.~I.~Uggerh\o j,
  Nucl.\ Instrum.\ Meth.\  B {\bf 234} (2005) 31.


\vspace*{-0.15cm} \bibitem{LHCC107} LHC Committee, minutes of the 107th meeting, CERN/LHCC 2011-010


\vspace*{-0.15cm} \bibitem{Baur00} A. Baurichter \etal, Nucl. Instr. Meth. B \textbf{164}, 27 (2000)


\vspace*{-0.15cm} \bibitem{meyer} W. Meyer, Habilitation thesis, Bonn, Germany (1988). 

\vspace*{-0.15cm} \bibitem{didelez} J.P. Didelez, 
     Nucl. Phys. News {\bf 4}, (1994) 10



\vspace*{-0.15cm} \bibitem{berlin} A. Berlin \etal, 
Procs. of 
     PSTP 2011, St Petersburg, Russia, 
p.  131.


\vspace*{-0.15cm} \bibitem{kohri} H. Kohri \etal, 
      Procs. of 
     PSTP 2011 St Petersburg, Russia, 
p.  142.


\vspace*{-0.15cm} \bibitem{solem} J. C. Solem, Nucl. Instr. and Meth. {\bf 117} (1974) 477 








\vspace*{-0.15cm} \bibitem{Sivers:1989cc}
  D.~W.~Sivers,
  Phys.\ Rev.\  {\bf D41 } (1990)  83.

\vspace*{-0.15cm} \bibitem{Brodsky:2002cx}
  S.~J.~Brodsky \etal, 
  Phys.\ Lett.\ B {\bf 530} (2002) 99;
%
  J.~C.~Collins,
  Phys.\ Lett.\ B {\bf 536} (2002) 43;
 %
  V.~Barone  \etal, 
  Prog.\ Part.\ Nucl.\ Phys.\  {\bf 65 } (2010)  267.


\vspace*{-0.15cm} \bibitem{Arsene:2008mi}
  I.~Arsene \etal\ [BRAHMS Coll.],
  Phys.\ Rev.\ Lett.\  {\bf 101 } (2008)  042001.

\vspace*{-0.15cm} \bibitem{Abelev:2008qb}
  B.~I.~Abelev \etal\  [STAR Coll.],
  Phys.\ Rev.\ Lett.\  {\bf 101} (2008) 222001

\vspace*{-0.15cm} \bibitem{Qiu:1991wg}
  J.~W.~Qiu, G.~F.~Sterman,
  Nucl.\ Phys.\  B {\bf 378 } (1992)  52.

\vspace*{-0.15cm} \bibitem{Schmidt:2005gv}
  I.~Schmidt, J.~Soffer, J.~-J.~Yang,
  Phys.\ Lett.\  {\bf B612 } (2005)  258.

\vspace*{-0.15cm} \bibitem{D'Alesio:2006fp}
  U.~D'Alesio, F.~Murgia,
  AIP Conf.\ Proc.\  {\bf 915 } (2007)  559.

\vspace*{-0.15cm} \bibitem{Bacchetta:2007sz}
  A.~Bacchetta \etal, 
  Phys.\ Rev.\ Lett.\  {\bf 99 } (2007)  212002.

\vspace*{-0.15cm} \bibitem{Qiu:2011ai}
  J.~W.~Qiu \etal, 
  Phys.\ Rev.\ Lett.\  {\bf 107} (2011) 062001


\vspace*{-0.15cm} \bibitem{Liu:2012vn}
  T.~Liu and B.~-Q.~Ma,
  Eur.\ Phys.\ J.\ C {\bf 72} (2012) 2037

\vspace*{-0.15cm} \bibitem{Boer:2012bt}
  D.~Boer and C.~Pisano,
  Phys.\ Rev.\ D {\bf 86} (2012) 094007


\end{thebibliography}
\end{document}